\documentclass{llncs}
\usepackage{graphicx}
\usepackage[utf8]{inputenc}
\usepackage[T1]{fontenc}
\usepackage{float}
\usepackage{hyperref}

\begin{document}

\title{Spatio-temporal profiling of public transport delays based on large scale vehicle positioning data from GPS in Wrocław}
\titlerunning{Spatio-temporal profiling of public transport delays}  
%
\author{Piotr Szymański\inst{1,3,4} \and Michał Żołnieruk\inst{1} \and Piotr Oleszczyk\inst{1} \and Igor Gisterek\inst{2,3} \and Tomasz Kajdanowicz\inst{1}}
\authorrunning{Piotr Szymański et al.} 

\institute{Department of Computational Intelligence, Faculty of Computer Science and Management, wyb. Stanisława Wyspiańskiego 27, 50-370 Wrocław.\\ Corresponding author: \email{piotr.szymanski@pwr.edu.pl}
\and
Department of Bridges and Railways, Faculty of Civil Engineering, Wrocław University of Science and Technology.\\ Corresponding author: \email{igor.gisterek@pwr.edu.pl}
\and
Society for the Beautification of the City of Wrocław, ul. Grabiszyńska 12/6, 53-502 Wrocław.
\and
illimites foundation, ul. Gajowicka, 53-530 Wrocław, Poland}

\maketitle              

\begin{abstract}
In recent years many studies of urban mobility based on large data sets have been published: most of them based on crowdsourced GPS data or smart-card data. We present, what is to our knowledge the first, exploration of public transport delay data harvested from a large-scale, official public transport positioning system, provided by the Wrocław Municipality. We evaluate the characteristics of delays between stops in relation to direction, time and delay variance of 1648 stop pairs from 15 mln delay reports. We construct a normalized feature matrix of likelihood of a given delay change happening at a given hour on the edge between two stops. We then calculate distances between such matrices using earth mover's distance and cluster them using hierarchical agglomerative clustering with Ward's linkage method. We obtain four profiles of delay changes in Wrocław: edges without impact on delay, edges likely to cause delay, edges likely to decrease delay and edges likely to strongly decrease delay (ex. when a public transport vehicle is speeding). We analyze the spatial and mode of transport properties of each cluster and provide insights into reasons of delay change patterns in each of the detected profiles.

\end{abstract}
\section{Introduction}
Why is my bus/tram late (again)? Every public transport user has asked this question while waiting at the bus/tram stop. With the rise of a number of inhabitants, cities struggle more and more to provide mobility services \cite{ceder2016public} that can prevent the rising number of cars in use, improve the quality of life and reduce the stress effect tardiness has on the population \cite{koslowsky2013commuting}. Understanding what causes public transport to be late is also important for the operator - with less delays the total traffic and transport performance can be improved with the same work and infrastructure costs. 

With this significance in mind we observe an important growth of intelligent transport solutions (ITS, \cite{mcqueen1999intelligent}) being deployed for public transport \cite{grant2014intelligent} - electronic ticket/smart card systems, cellphone-based mobility monitoring \cite{berlingerio2013allaboard}, cameras calculating the number of passengers on stops and in transit, gps-based location systems for automatic vehicle location (AVL) \cite{moreira2008travel} and many others. Karlaftis and Vlahogianni \cite{Karlaftis4} provide a broad summary of what has been done in this area before 2011.

Studies of urban mobility vary by the kind of data used, the mode (modes) of transport that they take into account and the motivation behind the study. Before the rise of larger data gathering systems scientists would collect information from personal travel journals provided by a selected group of study participants. Cell phone infrastructure availability allowed monitoring of macroscale mobility phenomena in urban areas \cite{csaji2013exploring}. Electronic ticketing systems and bike rental systems - the check-in - check-out data tables were evaluated. The rise of widespread GPS usage has brought large-scale data sources for all modes of transport - either from built-in devices in vehicles or from travelers' smartphones, while also joining them with other data \cite{giannotti2011unveiling}. 

Public transport analysis of smart-card and similar ticketing systems concentrated on understanding customer behaviour and mobility.  Morency et. al. \cite{morency2007measuring} use k-means (\cite{forgy-kmeans}\cite{macqueen-kmeans}\cite{steinhaus-kmeans}) clustering to create a mesoscopic perspective from 6 000 000 boarding information in Gatineau (Canada). The authors condition ticket type and times of usage on detected clusters. Each cluster obtained by clustering process  corresponded to different user profile like student or an elderly person. Similar works on passenger clusters analysis were done based on data from urban transport network of Rennes Metropole (France)\cite{Briand18}. Poussevin et. al. \cite{poussevin2014mining} perform nonnegative matrix factorization to detect clusters of activity patterns in Paris subway and explain usage behaviours. 

Having a GPS device in all vehicles gives possibility to mine a variety of information. Authors of \cite{Galba26} used buses gps locations from the period of one month to find spots with increased congestion (both k-means and DBSCAN \cite{ester1996density} algorithms were used on 12 000 data points). Stenneth et. al. \cite{stenneth2011transportation} detect stations and routes of public transport. Mendes-Moreira et. al. \cite{moreira2015improving} use GPS-based AVL system data to validate existing timetables. In Dalian City a custom variant of k-means algorithm was used to obtain locations of unique transit stops based on historical vehicle positions\cite{Xueying36}. This kind of systems can reduce costs of updating timetables with new data. Mazloumi et. al. \cite{mazloumi2009using} explore the GPS data from a bus route with spatiotemportal and weather considerations in mind.

We have not found reports which use large-scale GPS data containing information about punctuality of public transport vehicles such as the data used in this paper. However Rietveld et al. \cite{Bruinsma2} provide a noteworthy estimation of unreliability of public transport chains in Netherlands. The study is based on a medium-sized subset of transport operators, yet it's novelty lies in taking into account not only the delays but also the delay probability and set it against the user expectations and willingness to accept risk of tardiness.

In this paper we tackle the task of analyzing delays in a large-scale AVL data set collected from the official intelligent transport and positioning system of the public transport in the city Wrocław. Our goal is to provide insight into mesoscopic scale of public transport tardiness. We start with describing the data collection and preprocessing procedure in Section \ref{sec:data}, methodology of clustering in Section \ref{sec:meth} and the results with discussion in \ref{sec:results}.

\section{Data}
\label{sec:data}
In 2013 Wrocław's Municipal Transport Company (Miejskie Przedsiębiorstwo Komunikacyjne, MPK) provided the possibility to check location of each public transport vehicle's location. The map was shared on-line on company's website and is still available and actively used\footnote{\url{http://pasazer.mpk.wroc.pl/jak-jezdzimy/mapa-pozycji-pojazdow}}. MPK also created a mobile application leveraging both this data and some additional - information about delays. Inside the iMPK\footnote{\url{https://play.google.com/store/apps/details?id=pl.wasko.android.mpk\&hl=pl}} mobile application, users are able to see the delay of each public transport vehicle in Wrocław. This data is available to the mobile app via a JSON API which has been documented after the original application had been reverse engineered\footnote{\url{https://niebezpiecznik.pl/post/reverse-engineering-aplikacji-mobilnej-krok-po-kroku/}}. In this work we use data collected from this JSON API. It is the same data that the users of the official MPK application saw on their phones when waiting for the bus/tram.

\begin{table}
\caption{Description of data for each vehicle}
\label{table_data_format}
\begin{center}
\renewcommand{\arraystretch}{1.4}
\setlength\tabcolsep{3pt}
\begin{tabular}{llllll}
\hline\noalign{\smallskip}
Name      	& Data type & Example   & Description                                    &  \\
\noalign{\smallskip}
\hline
\noalign{\smallskip}
course\_id & \texttt{int}       & 8815994  & An id used for identifying a single course     &  \\
vehicle\_id & \texttt{int}       & 7355  & An id used for identifying a vehicle     &  \\
latitude   & \texttt{float}     & 51.092125 & A float with vehicle latitude position &  \\
longitude   & \texttt{float}     & 17.031378 & A float with vehicle longitude position &  \\
line\_no   & \texttt{string}     & 0L & A text identifying line which course is a part of &  \\
type   & \texttt{char}     & b & Information if a vehicle is a bus (b) or a tram (t) &  \\
direction   & \texttt{string}     & 12738 & A text describing the final stop of this course  &  \\
delay   & \texttt{int}     & 11000 & A number of milliseconds that the vehicle is delayed  &  \\
time   & \texttt{datetime}     & 2017-03-02 10:28:54 & A datetime with the moment of data snapshot &  \\
stop\_no   & \texttt{string}     & 21112 & A text with id of the last stop that the vehicle visited &  \\
\hline
\end{tabular}
\end{center}
\end{table}

For the purpose of this work data from Wrocław's Municipal Transport Company (MPK) was collected during a period of one month (February 2017). Every 10 seconds a snapshot of all vehicles positions was captured and saved with metadata (described in Table \ref{table_data_format}). Altogether 61 278 363 data points were collected between 1st and 28th of February 2017. This data was further reduced due to the fact that data points tend to be reported more than once, due to various reasons: cached results, the vehicle lost connection with GPS, etc. We have selected a subset of unique data points, taking the first minimum time stamp of a unique data point arrival. The selected subset consists of 25 755 313 data points (42\% of all data points). This data set however included both weekdays, weekends and night transit. In order to work on a coherent phenomenon, all of which differ in terms of the number of vehicles in the system, the trip frequency and also conditions. Delays occuring at night or on the weekend when traffic is reduced form different phenomena which should be analyzed separately. 

We have decided to perform our analysis only on weekdays, on which the daily mode of transport starts around 4:30 and slowly switches to night transit around 23:30. We have restrained the data to hours 6:00-20:00 on weekdays, so that we have encapsulated only the most standard daytime weekday public transport. As a result we have selected 15 409 910 data points, which encapsulate 59\% of unique data points and 25\% of all gathered data points.

\subsection{From vehicle positions to stops edges}
Our data contains locations of each vehicle with metadata. This format is not useful to us yet, because it would be difficult to aggregate it in a manner which would allow city analysts to draw valuable conclusions about delay patterns, in order to change this, the delay data was transformed. 

The value of feature \textit{delay} changes only when the value of feature \textit{last\_stop} does. It means that the course delay is calculated during the event of arriving to the next stop - it is the systems prognosis of the vehicles delay in arrival to the next stop. The important notion is that we are not interested in the delay \textit{per se}, as it might be more or less accurate in the end, but in the scale of changes of delays between stops. The fact that the delay changes only when the feature stop\_no changes is going to be the main assumption of the preprocessing process. For each course (a sequence of positions) algorithm checks whether the \textit{stop\_no} is different the the \textit{stop\_no} of the previous position. If this is the case, then we assign a difference of delays between these two positions to the edge between the previous value of \textit{stop\_no} and current one. The output is a list of delay changes which occurred while a bus or tram was traveling there. 

\subsection{Filtering based on official schedules}

Aside from vehicle positions data, we also use public transport schedule data from the General Transit Feed Specification\footnote{\url{https://developers.google.com/transit/gtfs/reference/}} for the city of Wrocław \footnote{\url{http://www.wroclaw.pl/open-data/}}. In our work we use two files from GTFS: \texttt{stops} which provide stops coordinates and \texttt{stop\_times} which we use to get all stop pairs being serviced by at least one line.

Having preprocessed our data, we stand with following structure:
\begin{itemize}
\item stop\_from: id of a stop which is the start of the edge
\item stop\_to: id of a stop which is the end of the edge
\item delay\_difference: difference of delays of vehicle when it was on \textit{stop\_to} and \textit{stop\_from}. It's how many miliseconds of delay the vehicle gained during the travel between \textit{stop\_to} and \textit{stop\_from}
\end{itemize}

It is now possible to compare edges from GTFS and edges which are a result of actions described in previous steps. We have inferred 19856 stop to stop edges from the AVL data, while the GTFS schedule allows 8252 edges (not including non-mandatory stops, which require a passengers action to stop, but including all night lines and special variants of lines). There are many reasons for this, most of the superflous edges stem from errors in reporting positions from AVLs or trips that had missed a stop. This can happen for various reasons: because the stop was not mandatory, because the vehicle's report was not delivered due to a GPRS connection error, because there was an unplanned detour (which happens relatively often). 

\begin{figure}[p]
\label{min_edges_maps}

\begin{tabular}{cc}
  \includegraphics[width=65mm]{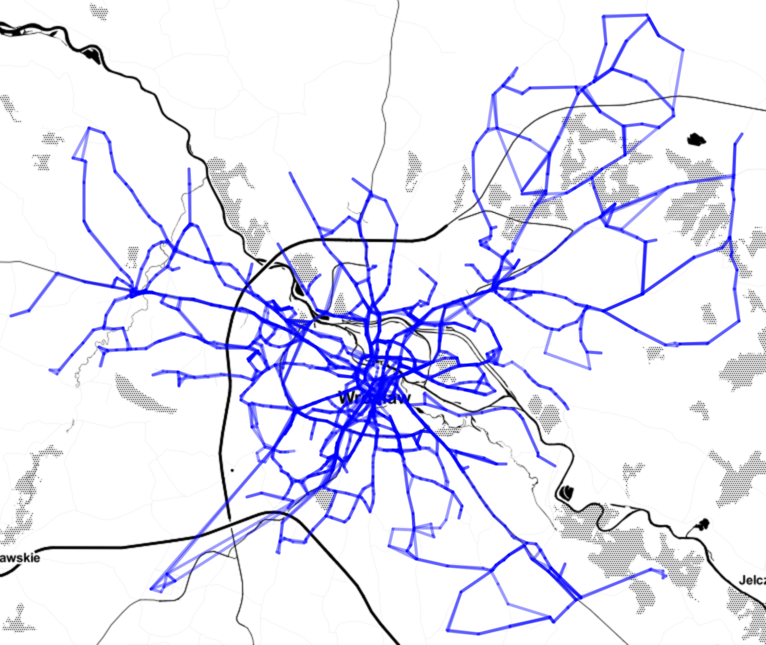} &   \includegraphics[width=65mm]{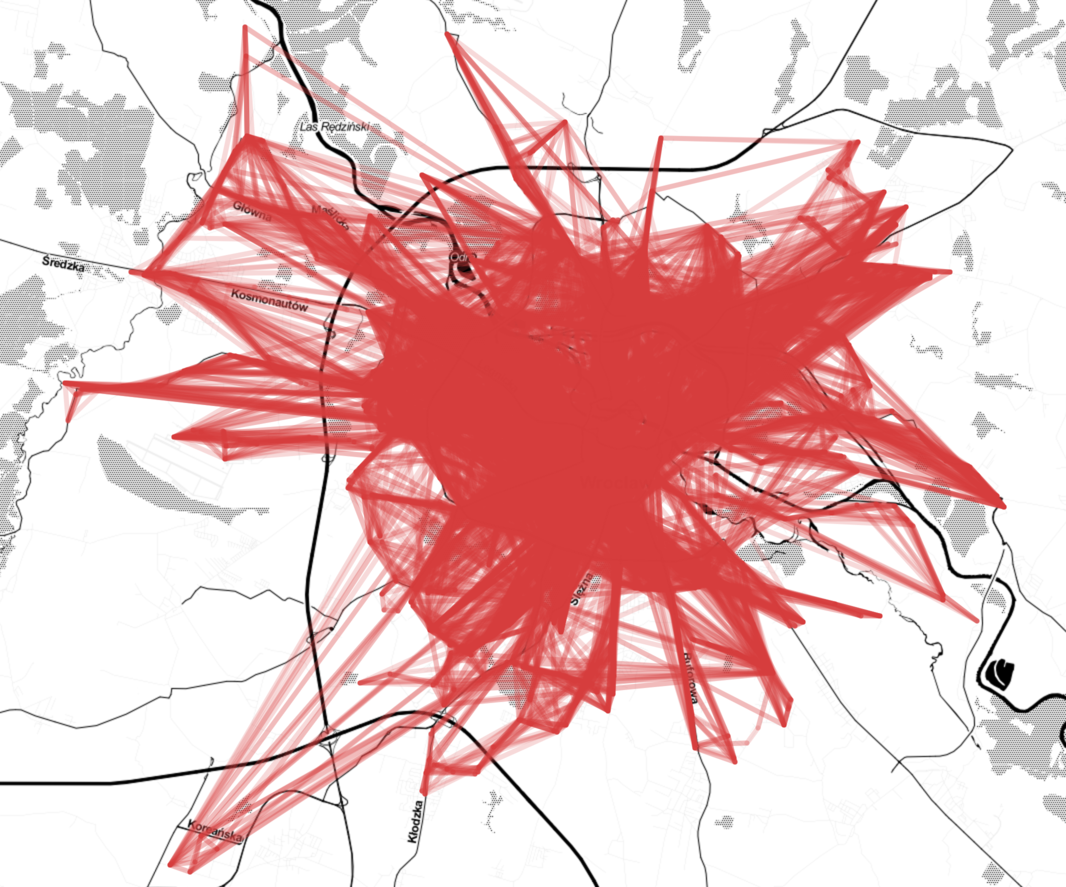} \\
(a) GTFS network & (b) D = 0  \\[6pt]
\includegraphics[width=65mm]{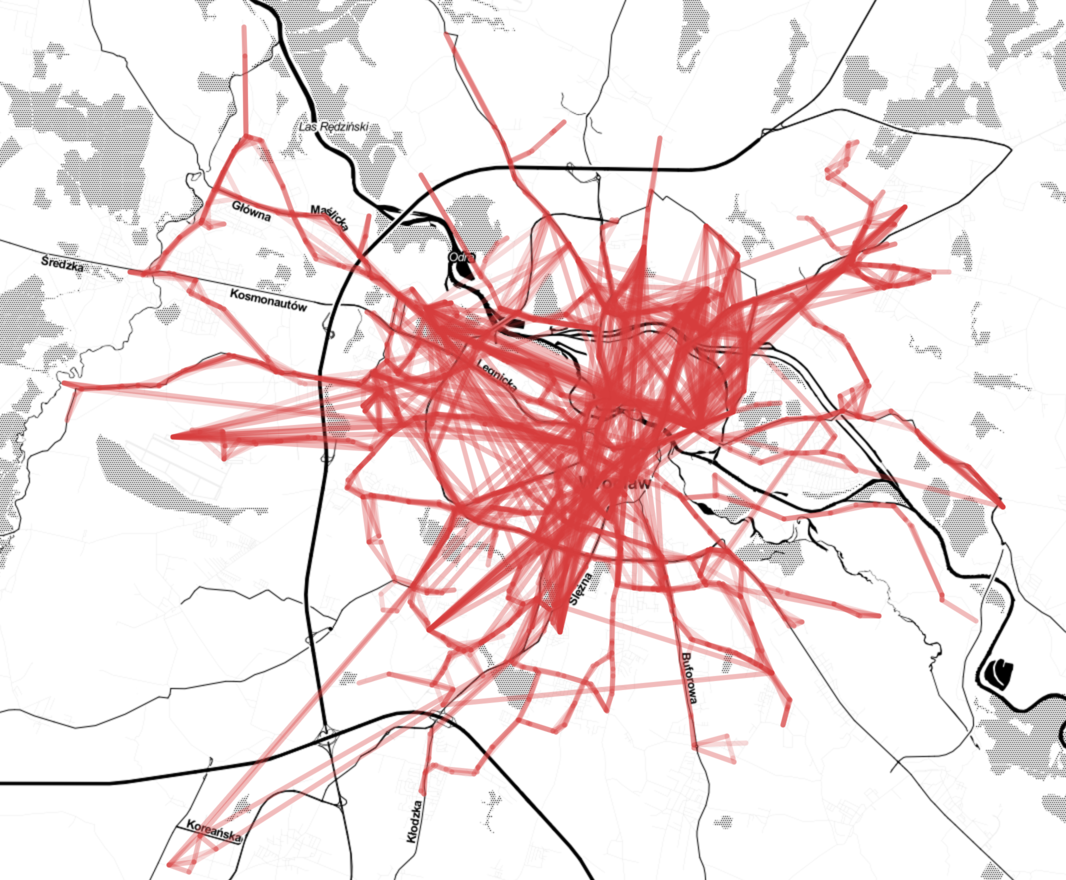} &    \includegraphics[width=65mm]{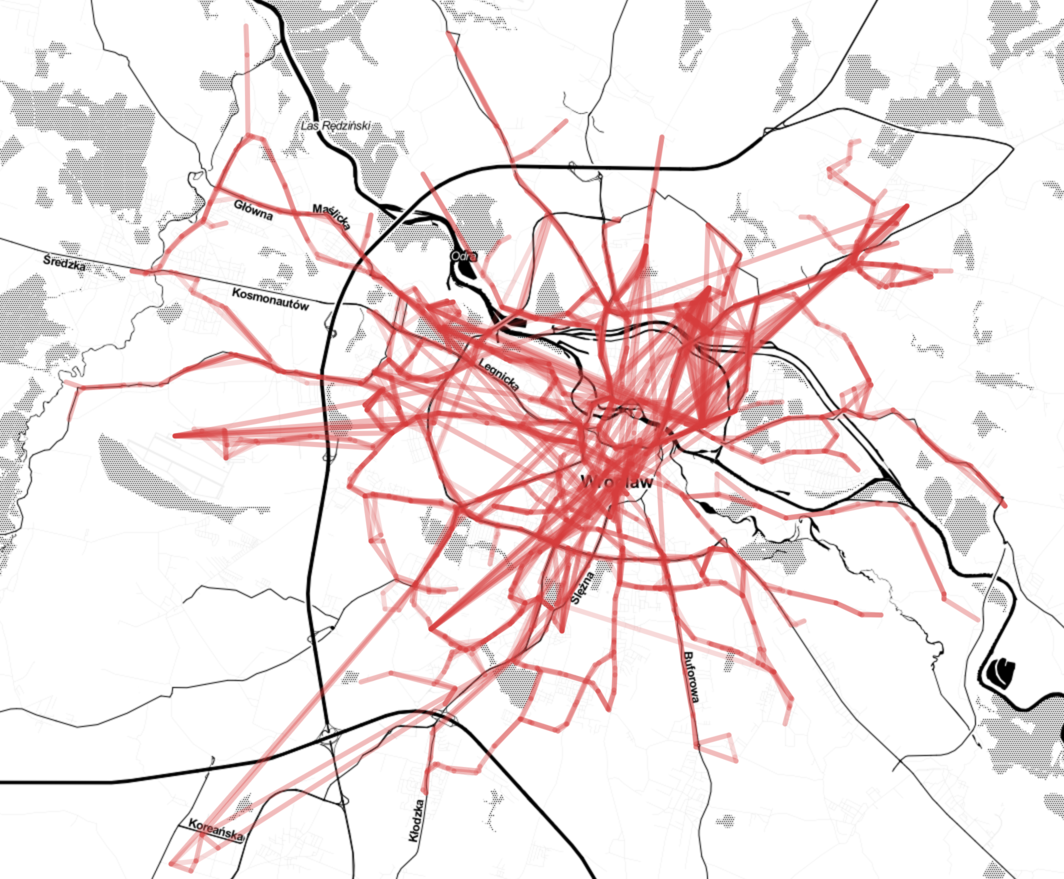} \\
 (c) D = 20 & (d) D = 50  \\[6pt]
 \includegraphics[width=65mm]{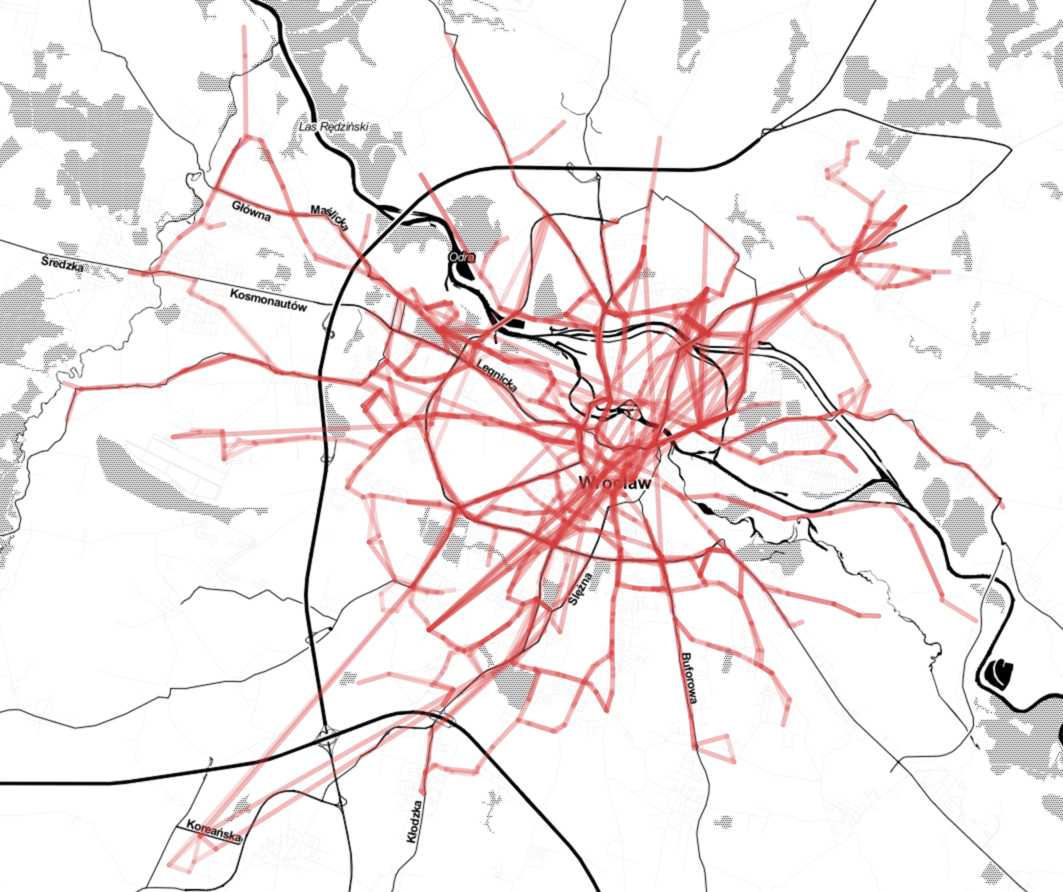} & 
 \includegraphics[width=65mm]{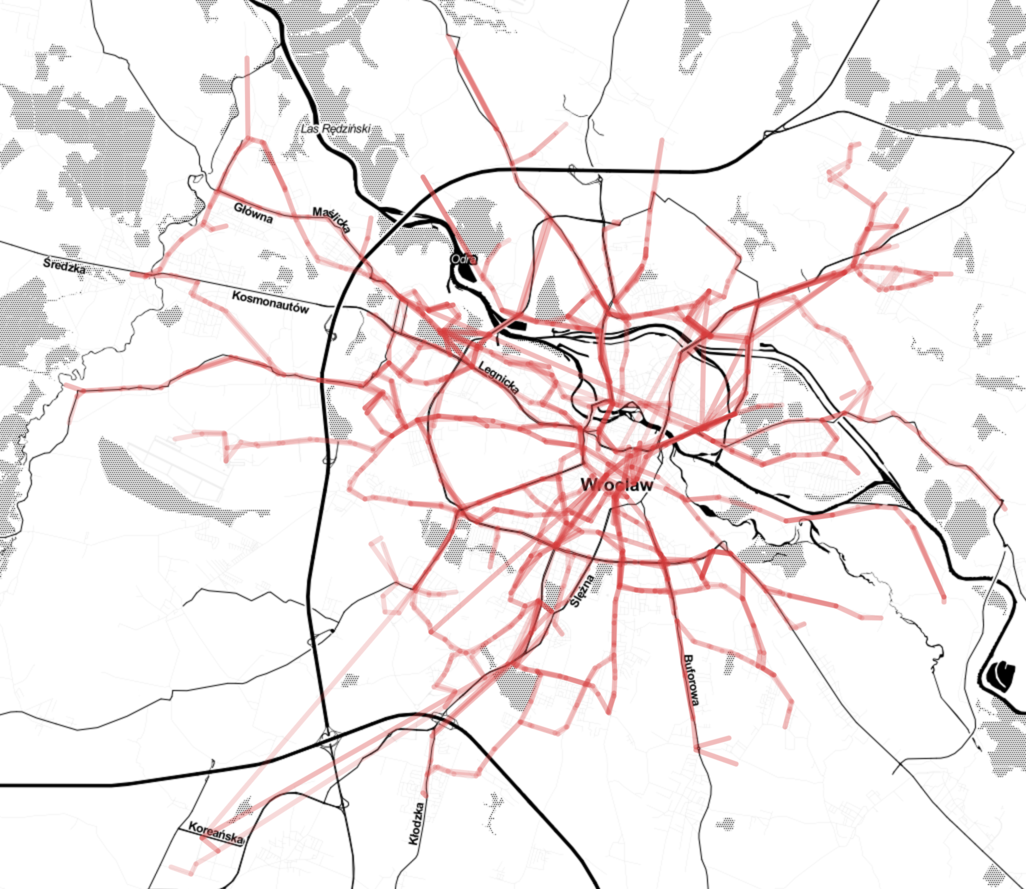} \\
(e) D = 100 & (f) D = 200  \\[6pt]
\end{tabular}

\caption{The chart above shows how the minimum count of delay changes on the edge impacts the number of made up edges and the coverage of GTFS edges}
\end{figure}

Luckily, these mistakes (edges which do not exist in real world) can be filtered quite easily. We have noticed that superflous edges contain much fewer data points than real ones. On other maps at Figure \ref{min_edges_maps} one can see how the graph of public transport changes when parameter D changes - an edge is shown only if it contains more delay changes than D. For D = 100 it can be observed that the density of network is comparable to the one from GTFS. We have selected D = 200 as the threshold, based on analysis of the outcome on the map and in histograms of delays. Such filtering that requires 200 data points per edge, i.e. at least 10 courses a day between 6:00 and 20:59, yielded 2108 edges. We have then filtered the edges that were consecutive mandatory stops in the schedule leaving 1648 edges. This also allowed us to remove special variants such as edges serving the purpose of vehicles joining in from/leaving to the the depot during the day.

\section{Methodology}
\label{sec:meth}
Each of 1648 edges which are a result of previous steps is represented by a list of delay changes of vehicles traveling on them. We want to perform cluster analysis to identify profiles of delay characteristics of public transport network. One can expect different groups of edges to have similar delay characteristics.

For each delay change our data also contains a timestamp with information when it happened. Intuitively one can assume that delay changes may vary during different time of day. A perfect representation would take into account:
\begin{itemize}
\item two dimensions of data: both delay change value and time it occurred should be considered,
\item density of points: the chosen representation should recognize the density of delay changes of given value in given time, i.e. be resilient to the fact that for certain times of day we may have more points than in others due to differentiation in transit headways.
\end{itemize}

\begin{figure}[!t]
\label{fig:tutorial_parta}
\begin{tabular}{cc}
  \includegraphics[width=65mm]{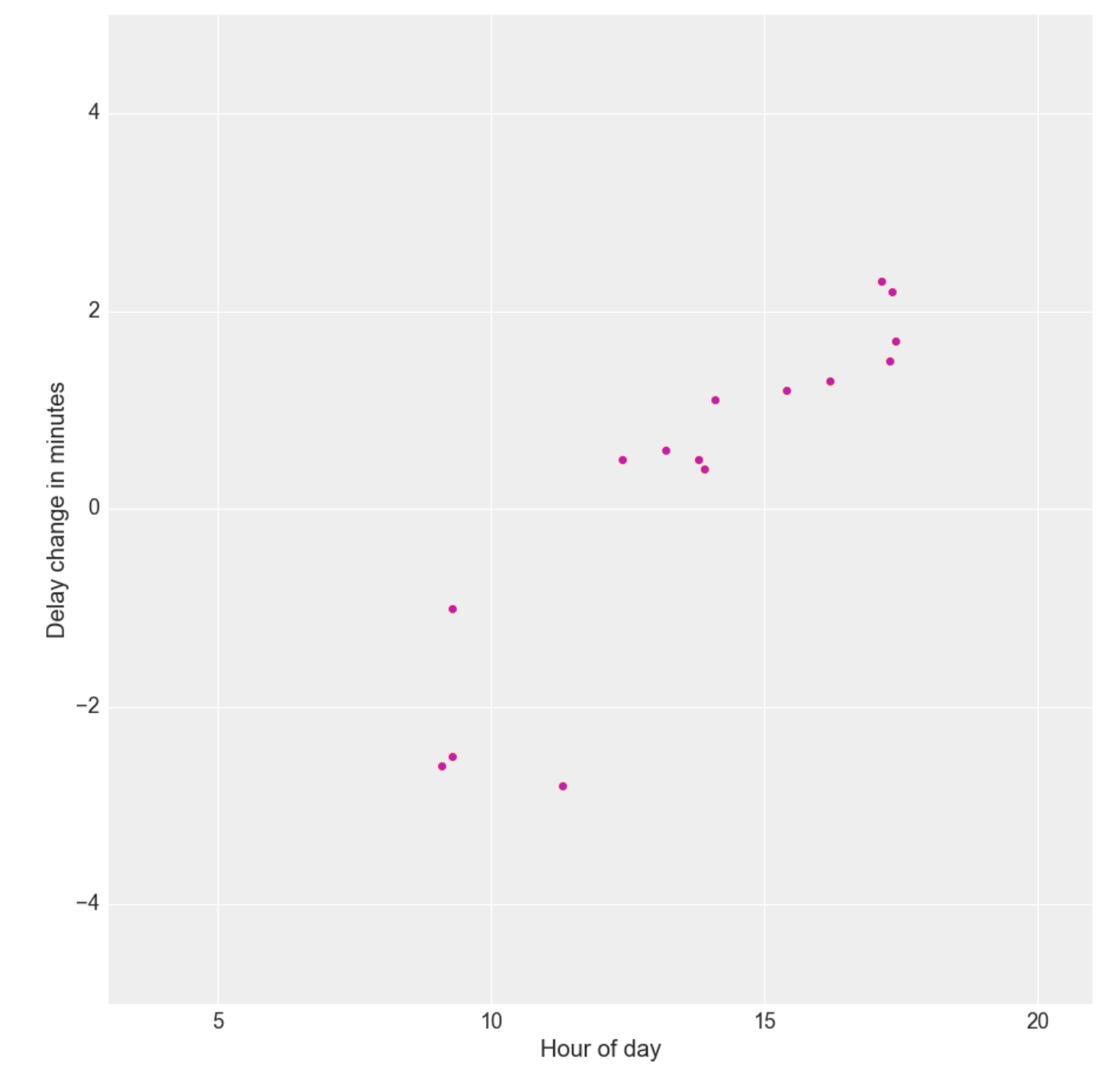} &   \includegraphics[width=65mm]{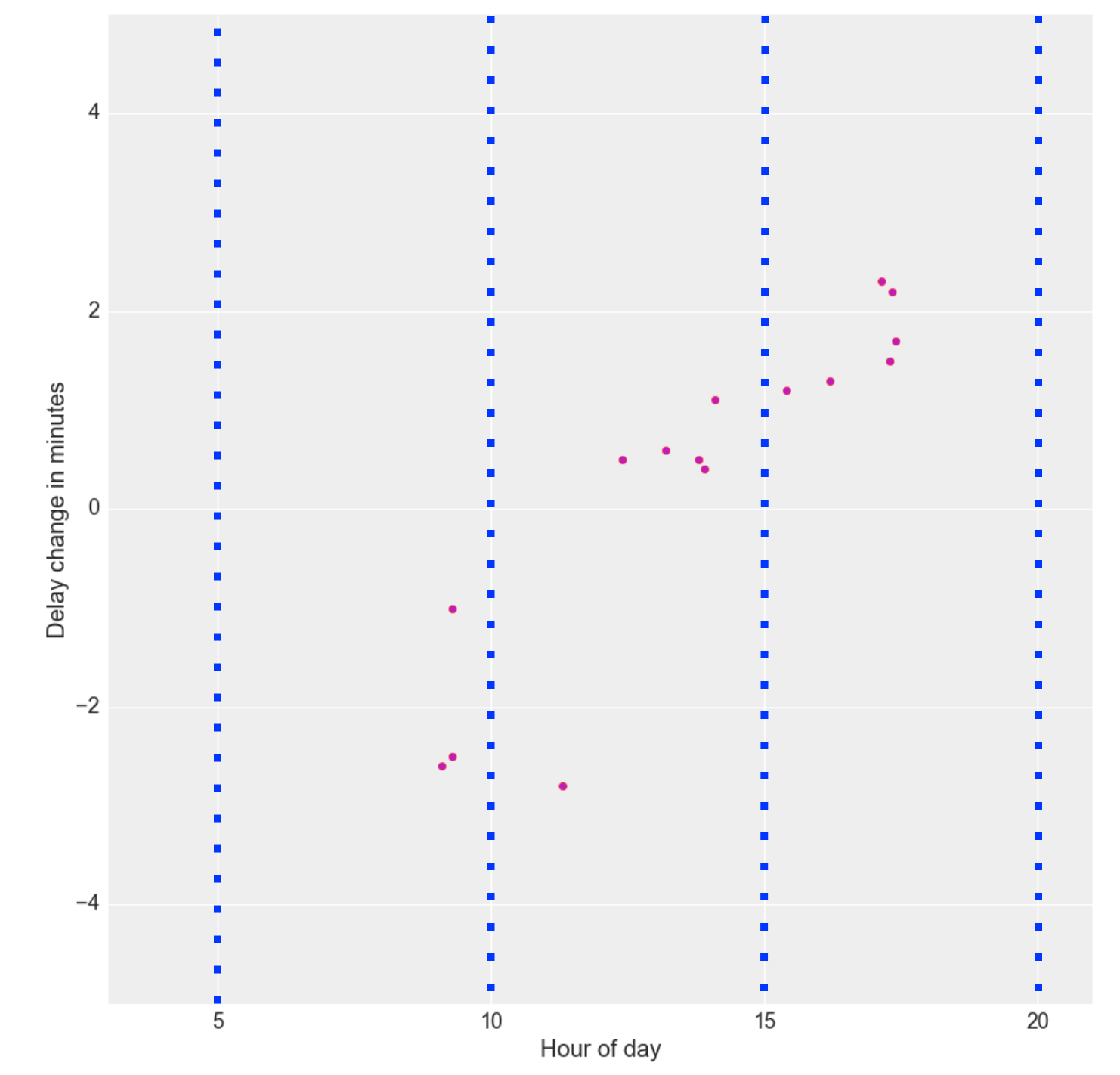} \\
(a) & (b) \\[6pt]
 \includegraphics[width=65mm]{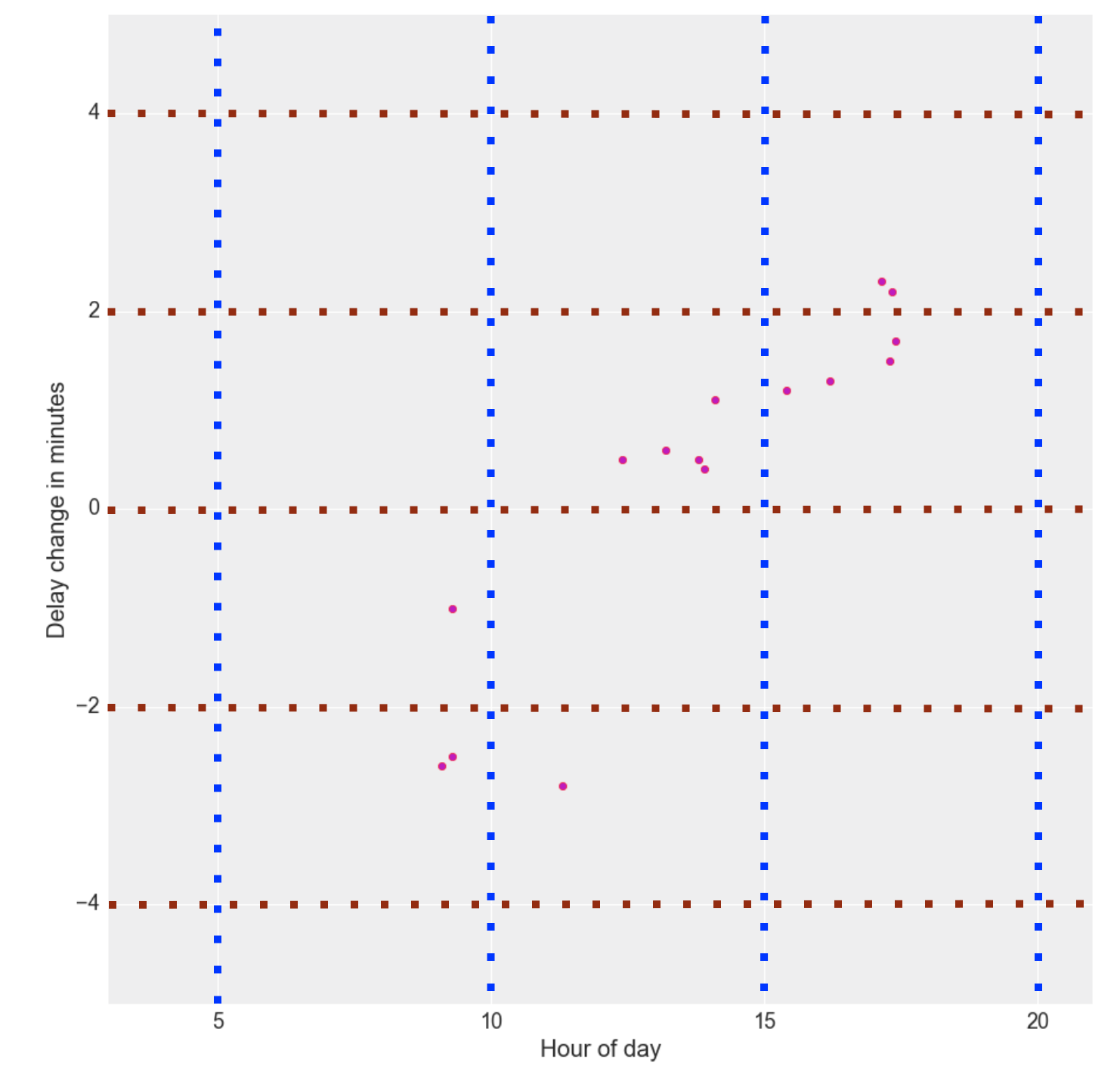} &   \includegraphics[width=65mm]{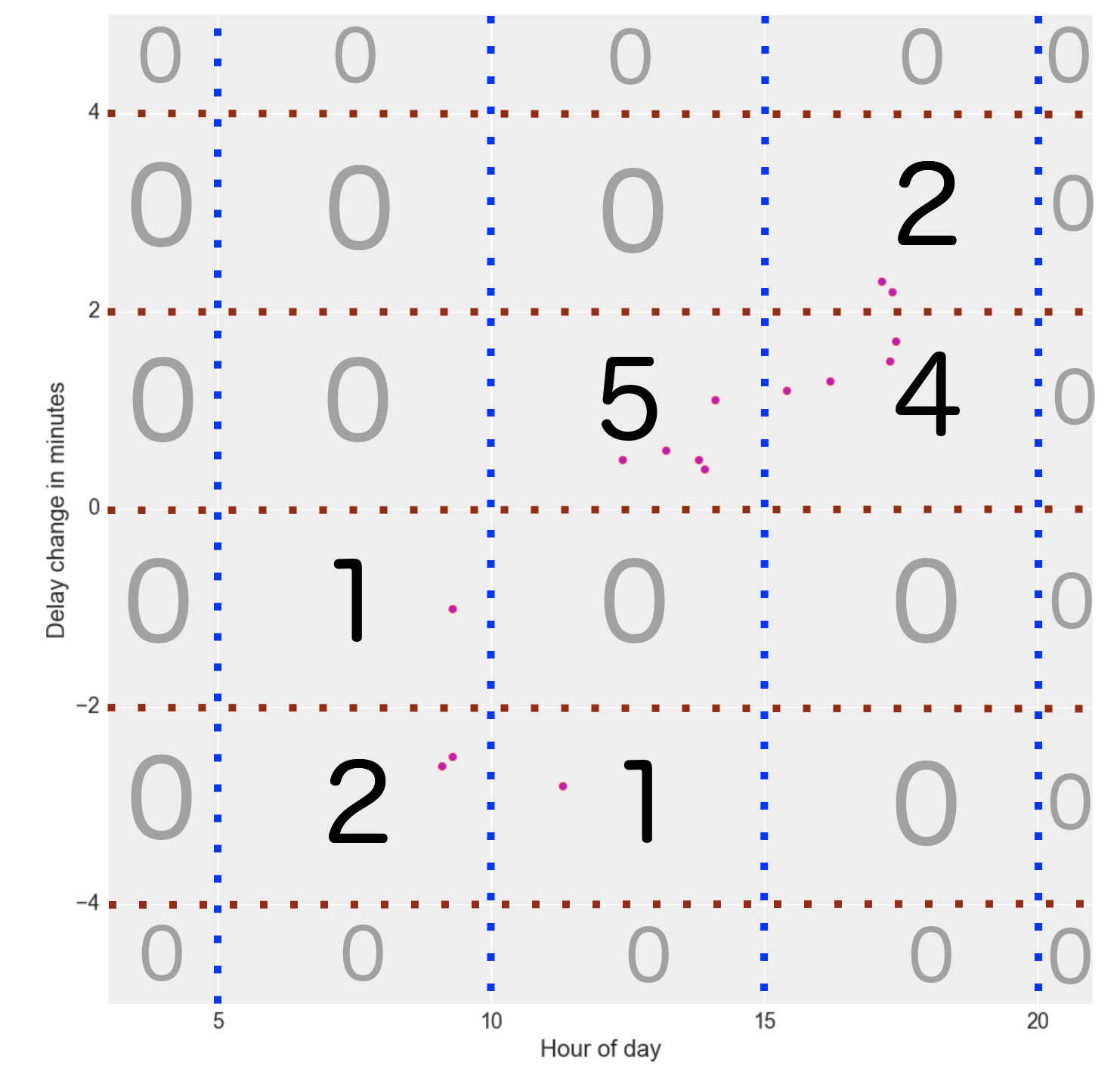} \\
(c) & (d) \\[6pt]
\end{tabular}
\caption{First part of processing for creating representation of edge delays suitable for further clustering}
\end{figure}

\subsection{Delay change distribution representation}
\label{sec:representation}
We propose a representation based on discretization of all delay changes points in both dimensions. Figure \ref{fig:tutorial_parta} shows first part of creating edge representation suitable for further clustering. Figure \ref{fig:tutorial_parta}a shows sample delay changes and hour they occurred plotted on a two dimensional chart. The next step is to decide the bins for time discretization. In the example the following bins were used (seen at Figure \ref{fig:tutorial_parta}b). The next step showed at Figure \ref{fig:tutorial_parta}d is to calculate total number of delay points in each bin (taking into account both delay change discretization and time discretization). 

\begin{figure}[!t]
\label{fig:tutorial_partb}
\begin{tabular}{cc}
  \includegraphics[width=65mm]{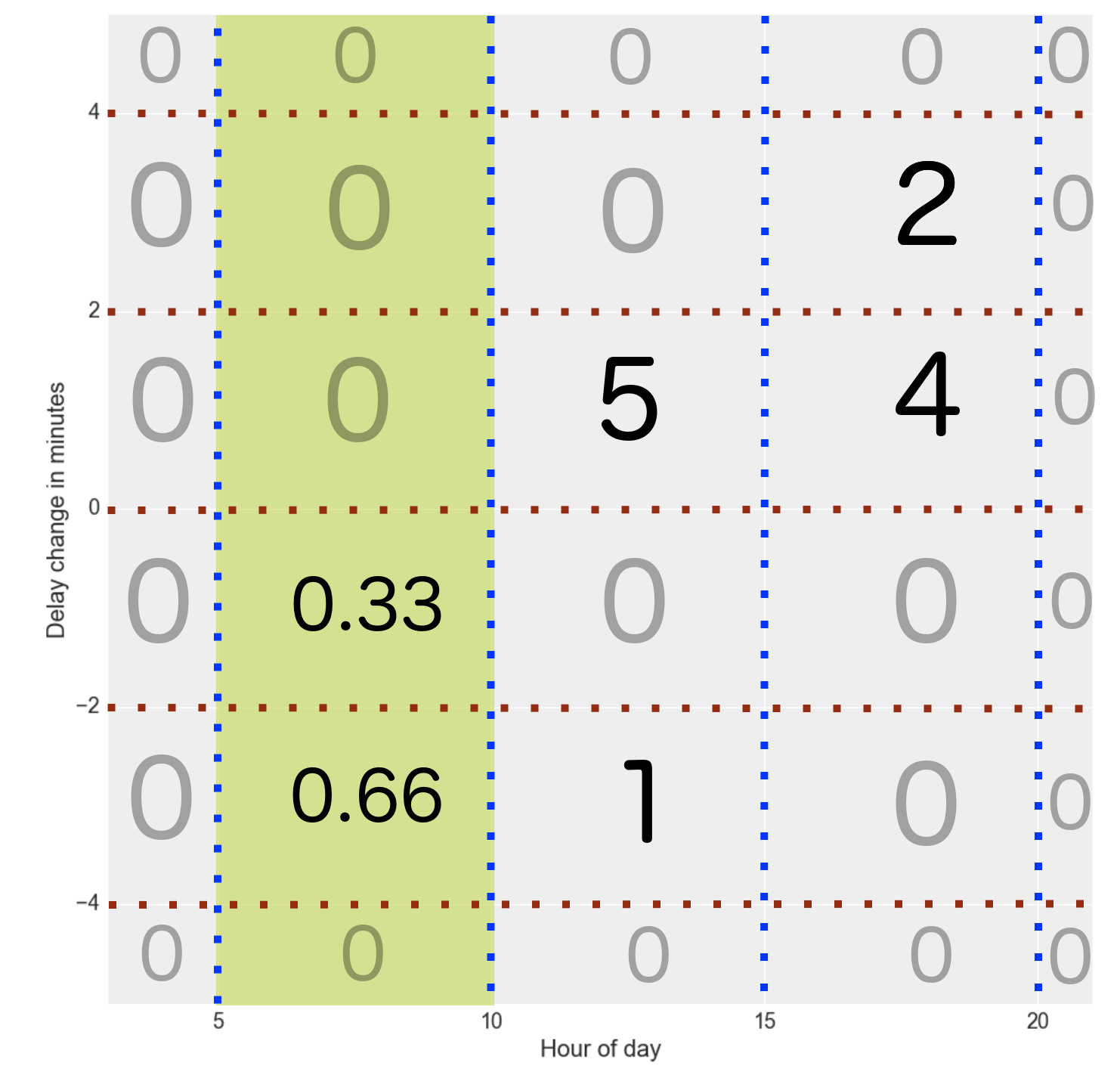} &   \includegraphics[width=65mm]{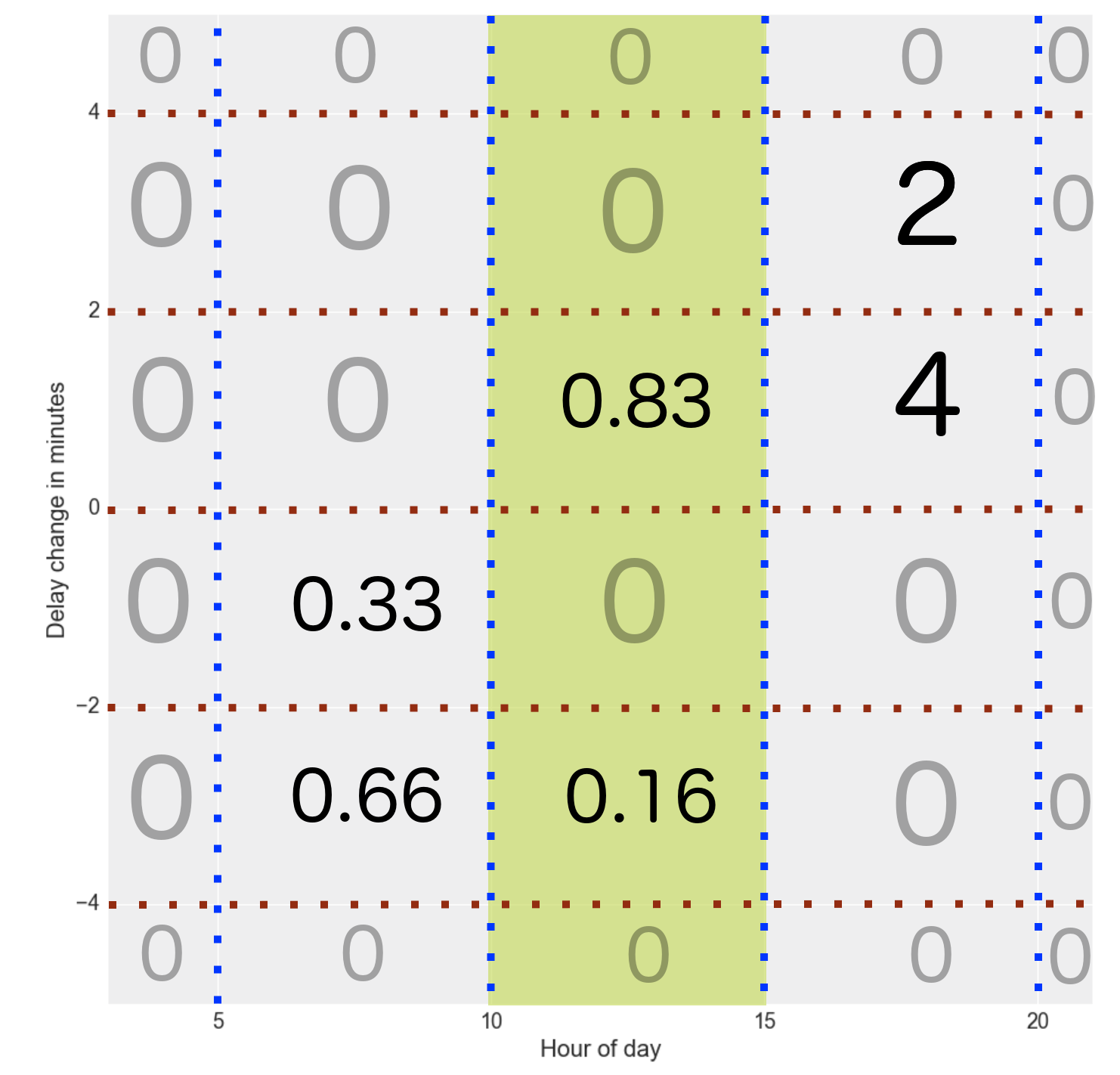} \\
(a) & (b) \\[6pt]
 \includegraphics[width=65mm]{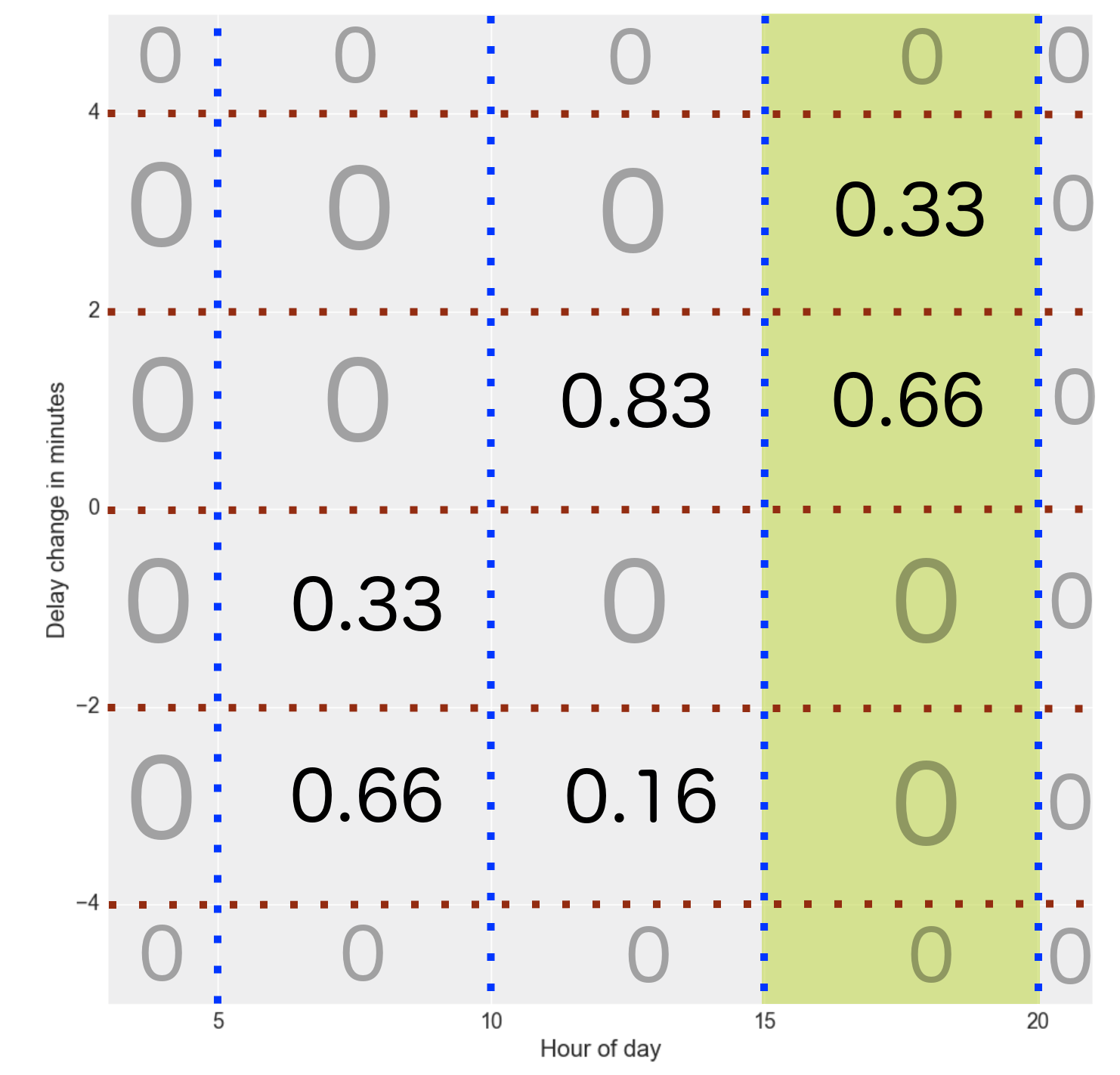} &   \includegraphics[width=65mm]{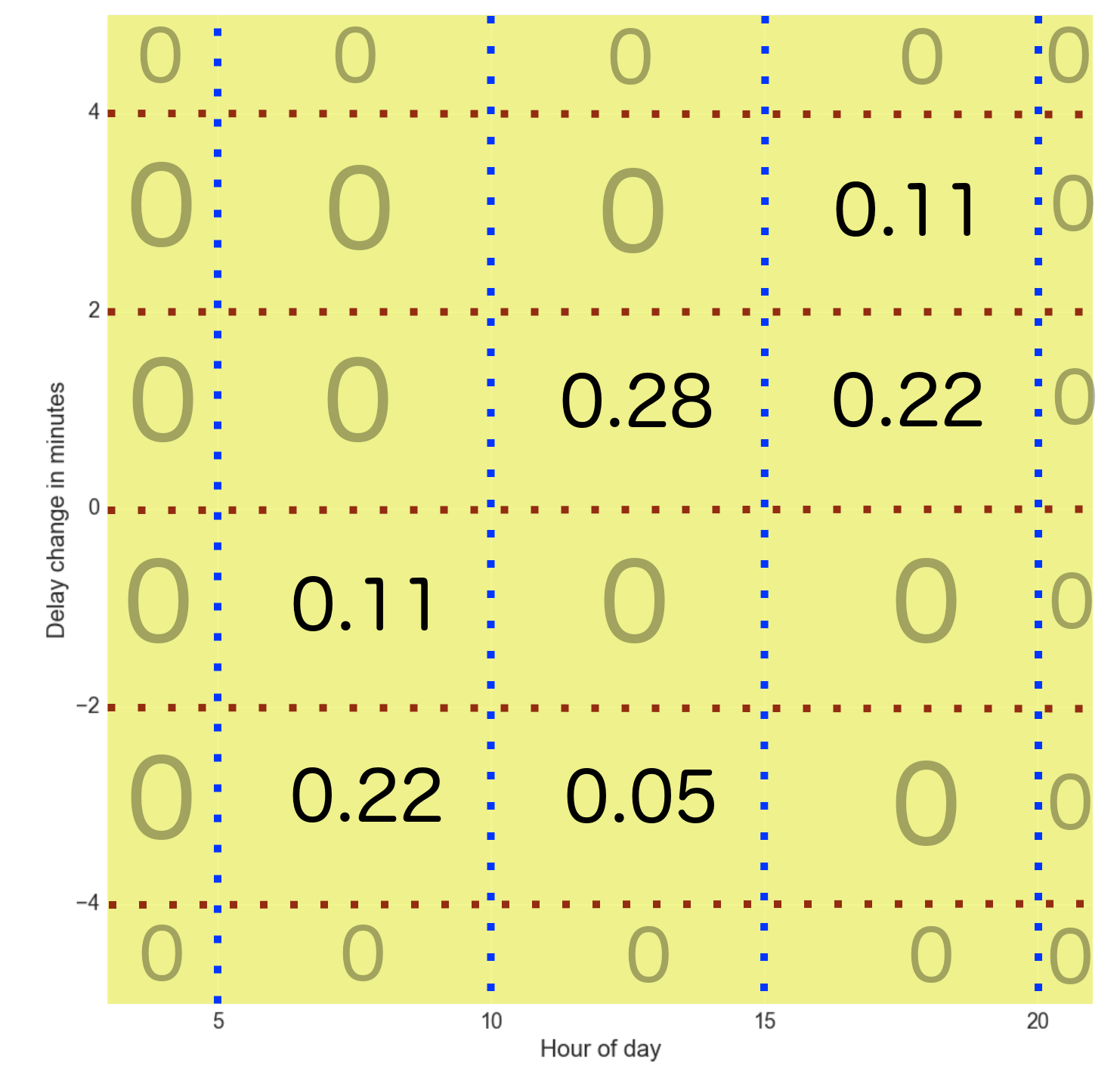} \\
(c) & (d) \\[6pt]
\end{tabular}
\caption{First part of processing for creating representation of edge delays suitable for further clustering}
\end{figure}

Then each bin for time is going to be considered separately and normalized (showed in Figure \ref{fig:tutorial_partb} (a)). For each cell in specific time bin the value is going to be normalized by dividing the value by the sum of values in whole column (time bin). This should be done for all rows as showed in Figure \ref{fig:tutorial_partb} (abc). This process of normalization guarantees that the number of delay points in whole time bin is going to be neglected during comparison with different time bin. The last step showed at Figure \ref{fig:tutorial_partb}d is to normalize the whole matrix - dividing each cell by the sum of all cells. Thanks to this process all the values of the edge representation sum to 1 and it is easier to analyze.

The edges which have no values in any of above time bins were deleted. The bins used for delays changes are as built in the manner: (-infinity, -30.5 min], then a bin of 5 minutes up till -5.5 - i.e. (-10.5,-5.5]. Next we use 1 minute long bins till -0.5, ex. (-5.5, -4.5]. The bin formed around 0, which we will refer to as punctual, on time, or no delay change bin is defined as (-0.5,0.5] bin, afterwards the rising delay bings are formed in a manner symmetric as the bins with negative changes in delay.

\subsection{Clustering}
There is no obvious choice for selecting a clustering method to obtain clusters of delay profiles in a public transport network. Many papers in the field use $k$-means, yet rarely do they profile justification for the choice of the method, or the choice of parameter. We decided not to use $k$-means as no well-grounded ways of selecting the parameter k without performing costly parameter estimation are available. Additionally our distance matrix between all edges will be precalculated and k-means algorithm relies on recalculating the mean representative of cluster, which wouldn't be possible without steps requiring heavy computations for each iteration of the clustering algorithm.

We have decided to use agglomerative clustering\cite{wardhierarchical} instead. This approach allows us to follow the dendrogram and understand each of the splits, allowing additional analysis and deeper understanding of the clustering process. Agglomerative clustering starts with assigning every edge to be a separate cluster. Then in each step it performs cluster merge between two nearest clusters. Clusters distances are maintained in a distance matrix, which at the beginning of the process is the same as the distance matrix between all objects (edges). After the merge of two closest clusters, the distance matrix has to be updated - algorithm removes distances between pairs containing merged clusters and recomputes the distance between new cluster and old clusters using a selected linkage metric. We have selected the Ward's method as the linkage metric as it connects those two clusters, when the connection will increase the sum of squares by the smallest value. It tends to create small clusters, but with elements really similar to each other. Such merges take place until all edges are merged into one cluster.

\subsection{Distance metric between edges}
In last chapter the data collected from ITS was processed into individual objects: edges starting and ending at given stops with delay changes distributed in time. Thanks to transforming them into discretized form, a matrix, it is now easier to compare two edges. The matrix has the same dimensions for all edges, which is going to make distance calculations simpler.

It is crucial for any clustering to choose a correct and domain fitting distance metric for comparing two objects. In our domain the distance metric should satisfy few requirements:
\begin{itemize}
\item it should reflect the difference in cell values
\item it should take into account the distance between cells which values are being compared
\item it should reflect the distance in a way that allows us to consider small variances of data as noise - really similar objects should be actually close to each other 
\end{itemize}

\subsubsection{Earth mover's distance (EMD)}
The concept of earth mover's distance was first proposed in 1781 as a part of transportation theory\cite{monge1781memoire}. It gained popularity as a distance metric for images. In 1941 earth mover's distance was again used a part of transportation theory by Hitchcock\cite{HitchcockEMD}. In his work he defines a problem of supplying amounts of a product from $m$ factories to $n$ cities. If one has to transport $x_{ij}$  tons of product from specific factory \textit{i} to specific city \textit{j}, which cost of is defined as $a_{ij}$, the total cost of all such operations would be:

$$y = \sum_{i=1}^{m}\sum_{j=1}^{n} x_{ij} a_{ij}$$

Stating the problem like this can make the distance we get using above equation proportional to the work one needs to perform to accomplish a task. As the name of EMD suggest, the source metaphor for this metric can be found in piles of earth. This distance is proportional to the amount of work that needs to be performed to transform one distribution (one field with piles of earth) into another (different field with different piles of earth). In our case the total amount of \textit{earth} is 1, since we normalized all of our bins at the end of processing edges to our representation for clustering. We used a python wrapper \textit{pyemd} for calculating earth mover's distance was used\cite{EMDpython}. It is based on the work done by Ofir Pele and Michael Werman\cite{pele2008}\cite{pele2009}.

Earth mover's distance considers all bins of our edge representation to be different piles of earth. It's goal is to find a minimal (in terms of performed work) way of transforming one edge representation to another. It does it using:

\begin{itemize}
\item values inside cells (popularity of given delay value at given time)
\item distances between cells
\end{itemize}

We described the first component in Section \ref{sec:representation}. The second one needs to be supplied into earth mover's distance algorithm in a form of a distance matrix. For each bin a distance to all other bins should be calculated. It should reflect the real distance between bins in both dimensions: delay and time changes. In this paper a simple work flow was used to calculate this distance matrix:
\begin{enumerate}
\item create a matrix with the same dimensions as the discretized representation we are going to calculate distances for 
\item in each cell, add coordinates with values of middles of both ranges it represents. If a cell is representing a time range between hours 5:00 and 6:00 and delay range between 3.5 minutes and 4.5 minutes, put the following coordinates to the cell: (5.5, 4.0). For the ranges with infinite values at the end, change them for to hard coded value for this step or take the non infinite part of range as the middle
\item if domains of your delay and time bins vary, consider scaling down or up appropriate coordinates. In this paper domain of delay changes was much broader than time changes, so it was divided by 4 for this step 
\item calculate standard euclidean distance matrix between all elements of created matrix
\end{enumerate}

\section{Results and Discussion}
\label{sec:results}

We obtained a dendrogram of clusters on the described data using the agglomerative clustering with Ward's method. We proceed with the analysis by the last 30 merges of the dendrogram, up until we reach a point where the clusters stop acting well enough as profile types and become too small and thus not general enough to form a typology. The smaller the cluster is, the more interesting it becomes as a source of insight concerning a concrete fragment of the city, yet it also begins to show extreme features of a subset of edges, which does not hold a generalization value.

We stopped after the third cut, which yielded 4 clusters of size 836, 262, 83 and 467 edges. Further cuts either did not provide generalized knowledge (ex. the last cluster would provide subtypes of being more likely to decrease delays by a larger number of minutes, but not change the overall profile of decreasing delays) or were small clusters of just a couple edges forming an interesting special case from an applied point of view but not viable in the scope of this paper.

\begin{figure}[H]
\centering\includegraphics[width=1\linewidth]{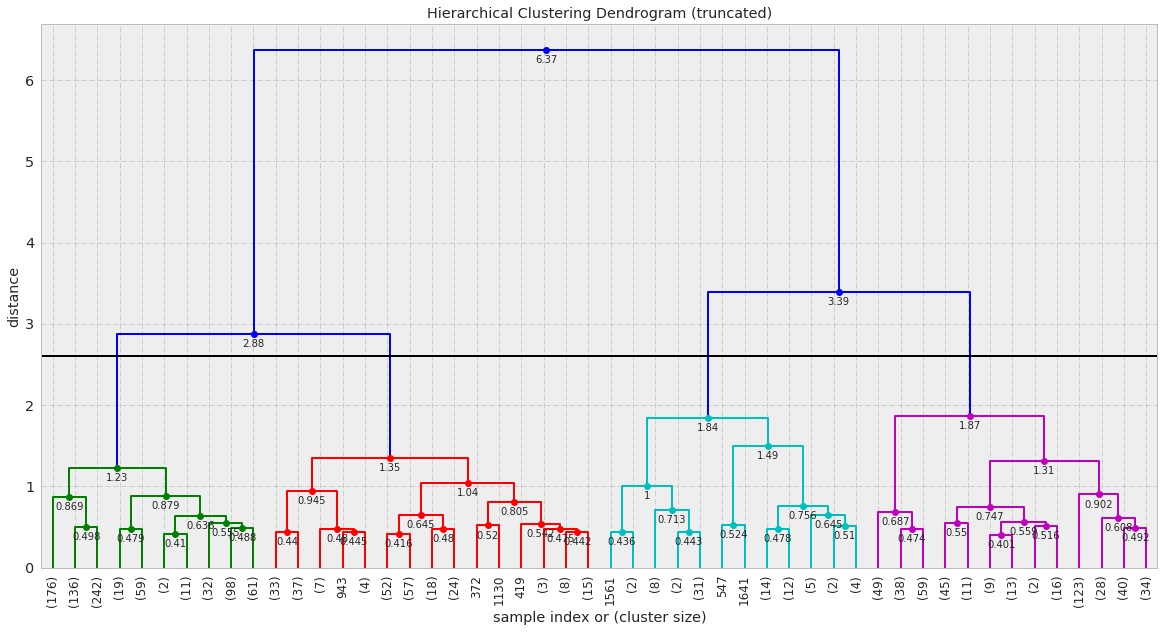}
\caption{Dendrograms of obtained clusters. Number at the bottom, if in parenthesis around denotes the size of the cluster. If it's just a number, then it's a cluster with one element.}
\label{fig:emd_histogram}
\end{figure}

We will evaluate the clusters based on the the spatial distribution of edges in the cluster and the distribution of delays.

The first cut (we denote it as 1.1, following the scheme cut:cluster number), shows two clusters out of which the one on the left side (1098 edges) of the dendrogram concerns edges which are most likely not to increase delays: 60\%, increase delay: 26\%, decrease delay: 12\%. All in all the cluster's most significant delay changes spanned from: -2.5 to 3.5 minutes. On the other hand the second cluster (2.2, 550 samples) exhibited delay changes from -10.5 minutes to 3.5 minutes. The likelihood of delay decrease exhibited was 57\%, increase 10\%, no change 31\%. Both clusters did not provide significant divisions to allow insight from spatial distribution of edges per cluster. 

In the second cut, the second cluster from the first cut (cluster 1.2) splits into a cluster 1.2:2.2 with 83 samples which spans delay changes from -10.5 to 3.5 minutes and the most probable delay changes span from -3.5 to 0.5 minutes: no change (-0.5 to 0.5 min all together) 15\%, -1.5 to -0.5: 18\%, -2.5 to -1.5: 23\%, -3.5 to -2.5 19\%. Delay decrease of 5 to 10 minutes is likely to happen with 2\% chances. This is a cluster of edges likely to decrease delays by large values.

The second branch of the split is cluster 1.2:2.3 which consists of 467 samples with two delay changes of highest likelihood during the day (well concentrated in two rows in the feature matrix): no change: 30\%, -1.5 to -0.5 minutes: 40\% likelihood and -2.5 to -1.5 minutes 11\%. Larger delay changes are less likely, with changes happening outside of the 0.5 to -3.5 less likely to happen than 1\% of the trips for each case delay interval. This is a cluster likely to decrease delays by small values.

As this side of the dendrogram does not change further, we end up with clusters denoted as: 1.2:2.2:3.3 (in short cluster 3) and 1.2:2.3:3.4 (cluster 4). 

In the third cut the cluster 1.1 (i.e. the left subtree) splits into two clusters: 1.1:3.1 and 1.1:3.2. The first one groups 836 edges that maintain the highest percentage of data points of being on time: 67\%. The delay probabilities in both decrease and increase are symmetric, the most probable cases are: -1.5 to -0.5 minutes: 12\% and 0.5 to 1.5 minutes: 16\%. This cluster shares the largest resemblence to cluster 1.1. We can see that the closer to the left we are on the dendrogram, the more punctual the edges are. 

The subcluster 1.1:3.2 consists of 262 edges that are more likely to gain delay though the likelihood of being on time remains high: increase of delay between: 0.5 to 3.5: 51\%, being on time 37\%. These cluster represents a mirror phenonemon of cluster 1.2:2.3:3.4. These clusters are denoted: 1.1:2.1:3.1:4.1 (cluster 1) and 1.1:2.1:3.2:4.2 (cluster 2).

\begin{figure}[h]
\label{fig:cluster_dist}
\begin{tabular}{cc}
  \includegraphics[width=65mm]{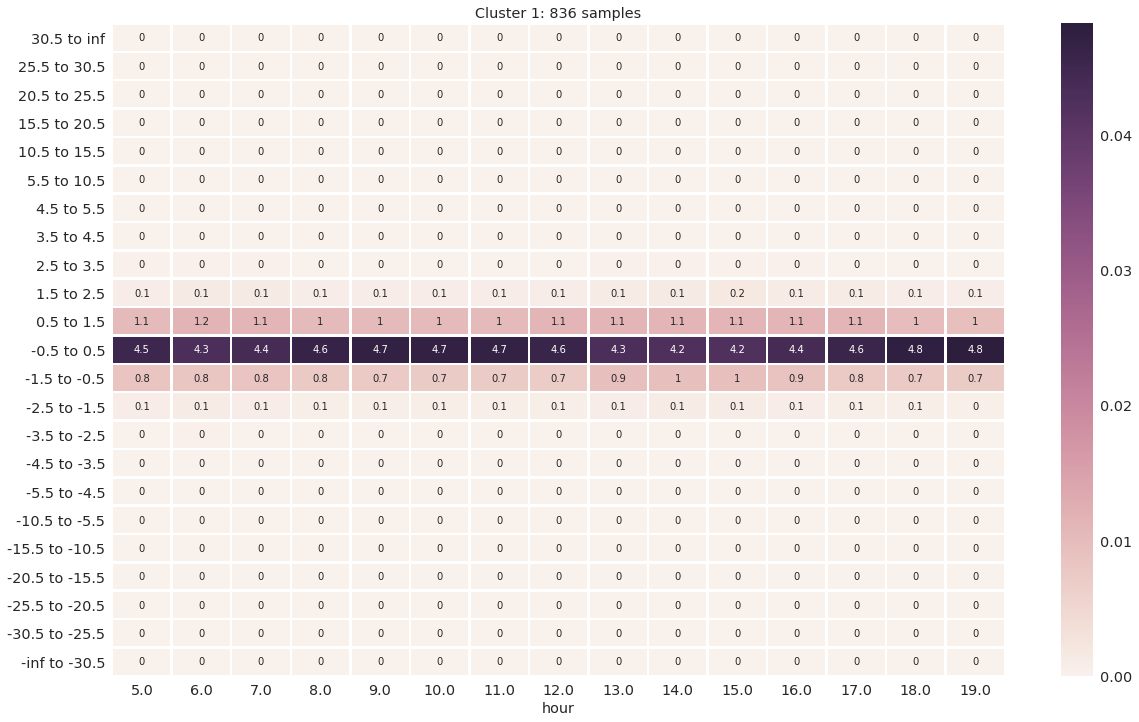} &   \includegraphics[width=65mm]{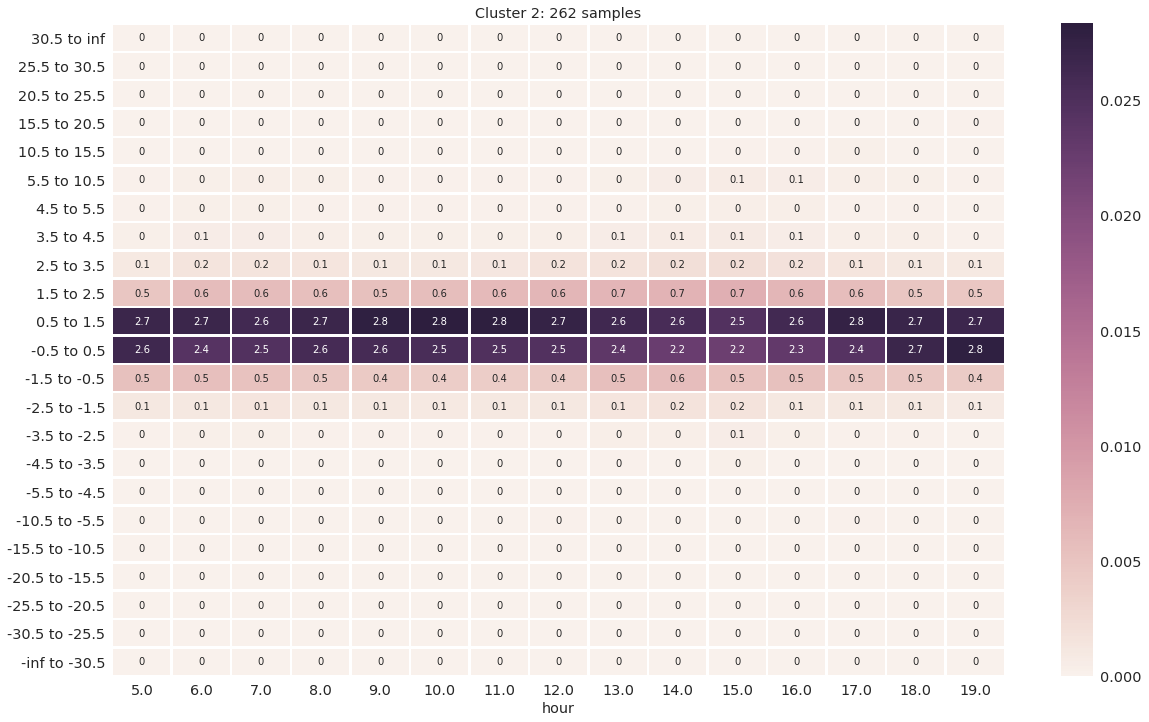} \\
(cluster 1) & (cluster 2) \\[6pt]
 \includegraphics[width=65mm]{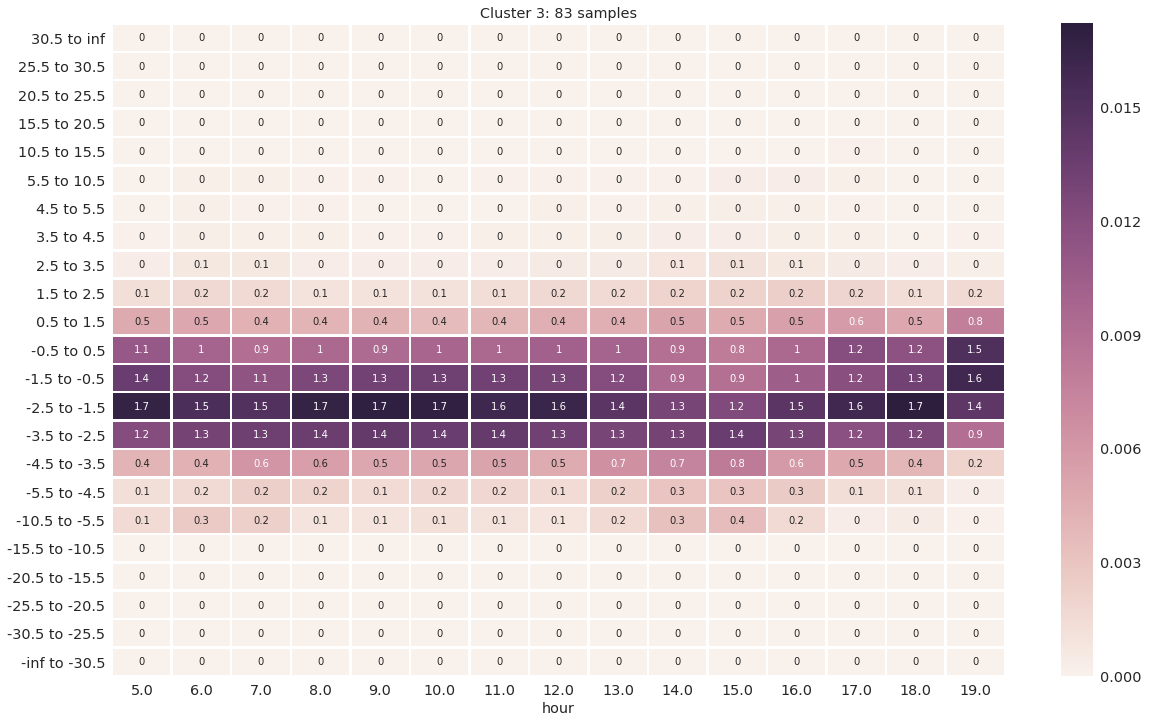} &   \includegraphics[width=65mm]{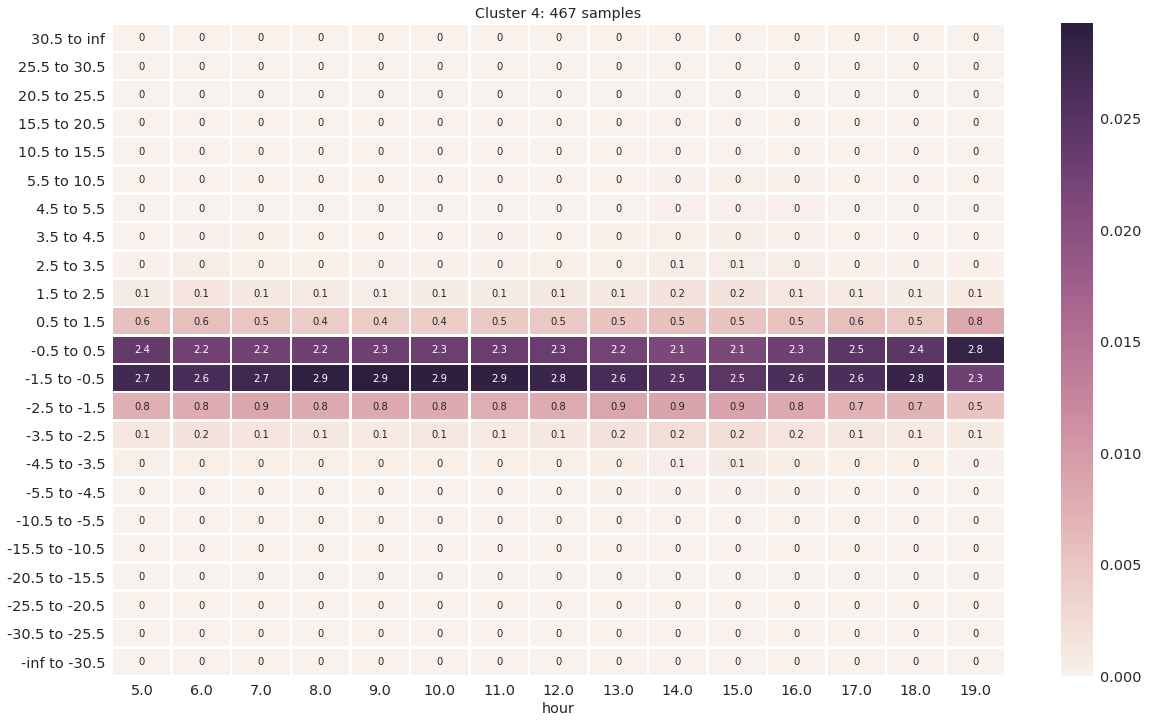} \\
(cluster 3) & (cluster 4) \\[6pt]
\end{tabular}
\caption{Matrix representation of delay likelihoods per hour aggregated in each of the clusters}
\end{figure}

Delay likelihoods in the clusters are provided in Figure \ref{fig:cluster_dist}.

These four clusters form the four basic archetypes of stop edges in Wrocław:
\begin{enumerate}
\item not impacting the delay significantly
\item likely to cause increase of delay
\item likely to cause strong decrease of delay (speeding, trying to make up time for delays)
\item likely to cause decrease of delay
\end{enumerate}

It is also interesting to note that from a systematic point of view the delays in public transport in Wrocław come from accumulating small delays over multiple edges rather than from single edges causing vast delays.

\begin{figure}[h]
\centering
\includegraphics[width=1\linewidth]{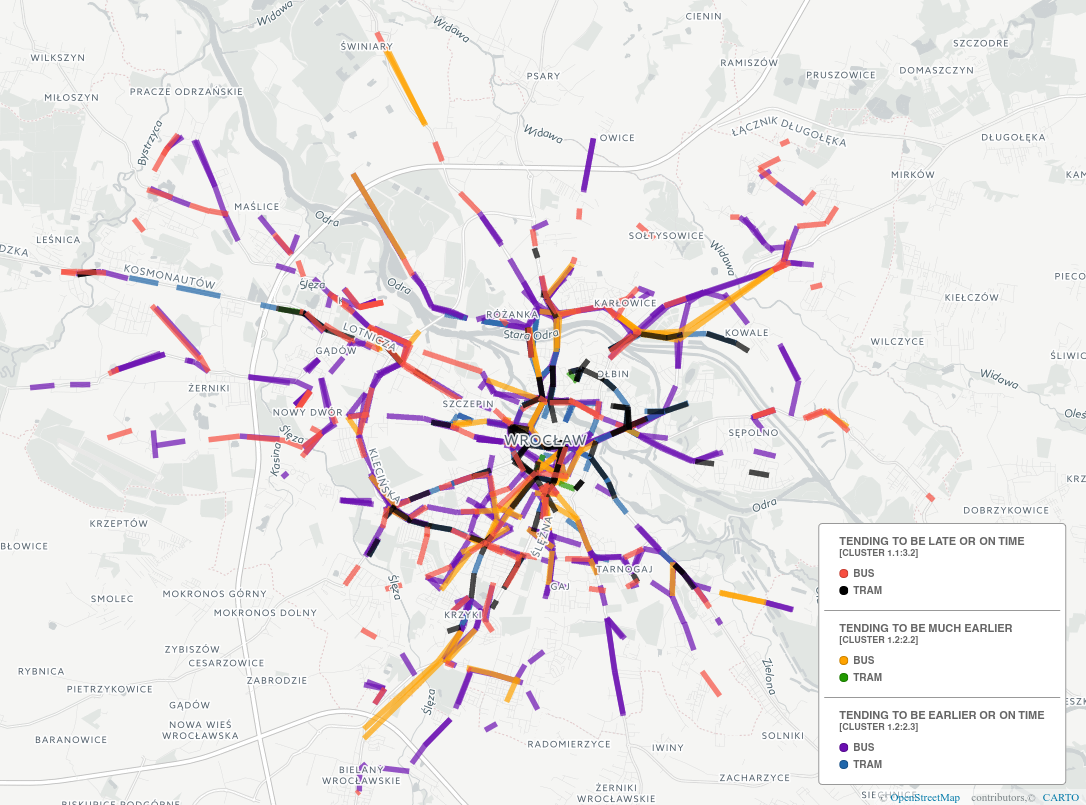}
\caption{The three delay impacting clusters mapped together.\href{http://niedakh.carto.com/viz/c3aa7d59-e72b-4514-b722-4b0d25ad968d/public\_map}{Interactive version is also available.}}

\label{fig:threecuts}
\end{figure}

Mapping the results allows us to see certain interesting phenomena (see Figure \ref{fig:threecuts}) such as: black tram delay increasing edges in the city center, yellow bus delay decreasing (possible speeding) edges on the outskirts of the city, bus delay edges on the crowded innercity ring which does not have separate bus lanes. Yet while we have already found clusters that separate some of the extreme values on the right side of the dendrogram, i.e. the edges that reduce delays by more than the next minute - we did not find such clusters on the delay gathering side.

It can be clearly seen that in afternoon traffic tramways are almost entirely free from significant delays, and sections characterized by high delay gains or losses are occasional. Also, several short sections tend to decrease delays which makes travel time stable. For buses the situation is much different. Five significant and quite long sections suffer from unpredictability, tending to either increase or decrease delays. Few sections tend to increase delays, while about dozen spots scattered in midtown enable buses to fight delays.

\begin{figure}[h]
\centering\includegraphics[width=1\linewidth]{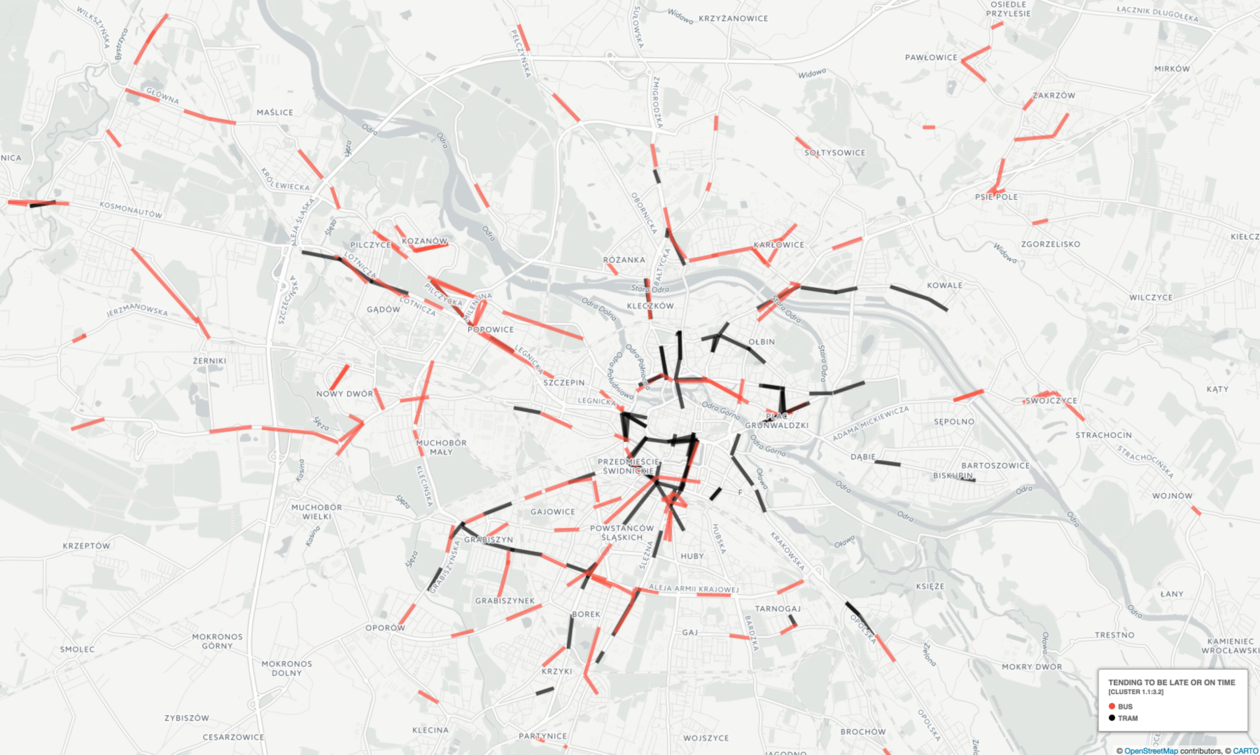}
\caption{Cluster 2 mapped separately, divided by mode of transport (t)ram/(b)us}
\label{fig:cl2_map}
\end{figure}

\subsubsection{Cluster 2: Likely to increase delay}
Concerning buses, majority of these sections is evenly distributed on the road network of the city, while the minority forms four radial corridors and one path on the southern bypass. It is mostly visible on long sections shared with heavy car traffic, showing further need of separating proper right of way. As for tramways, most of the cases forms a ring around the Old Town. Knowing that almost all of these cases use own right of way, a conclusion can be deducted that much improvement must be made in traffic signal programs.

\begin{figure}[h]
\centering\includegraphics[width=1\linewidth]{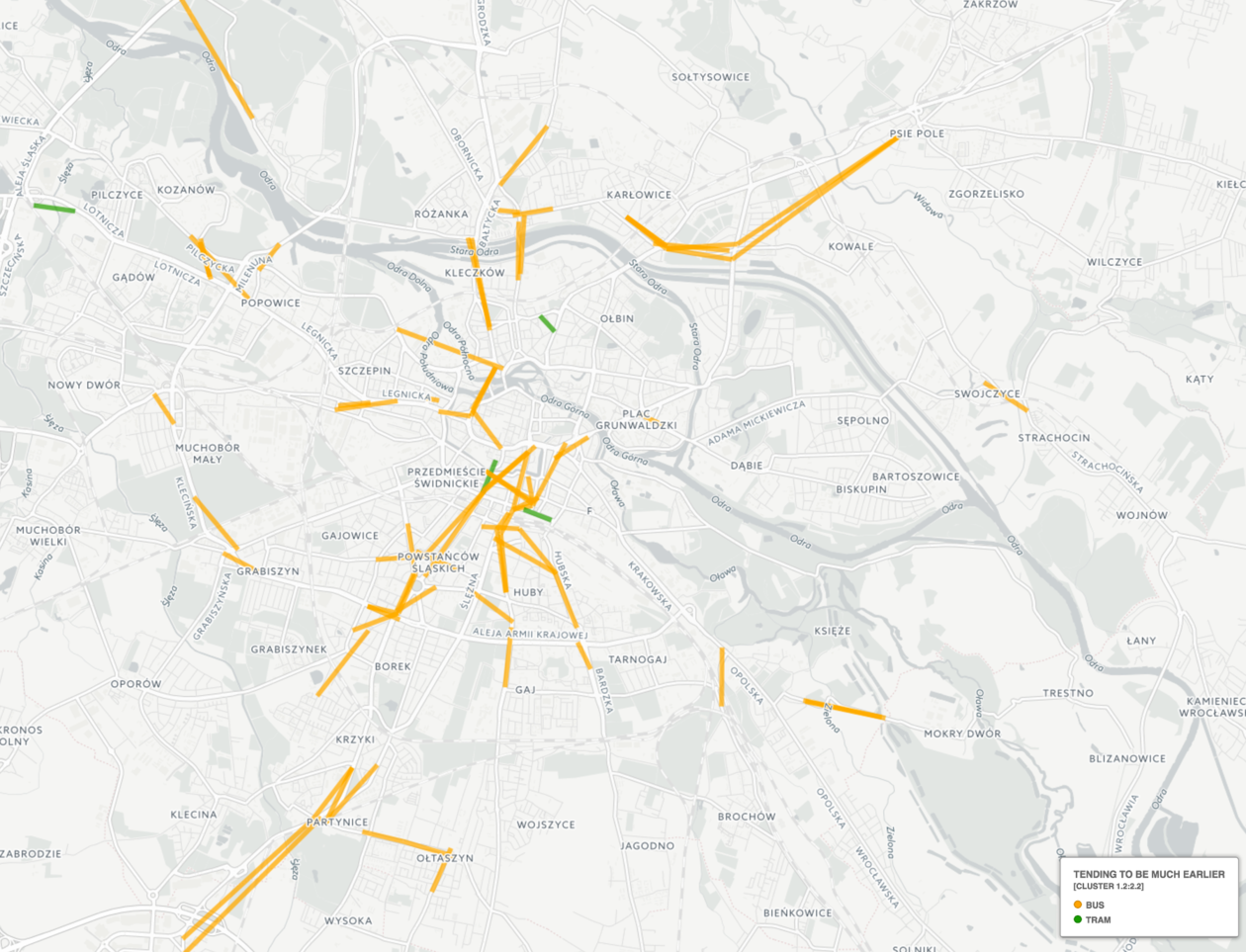}
\caption{Cluster 3 mapped separately, divided by mode of transport (t)ram/(b)us}
\label{fig:cl3_map}
\end{figure}

\subsubsection{Cluster 3: Likely to decrease delay}
This trend occurs in important radial bus corridors in southern and northern part of the city, and also partially on inner bypasses. It is most significant on long sections shared with heavy car traffic. Meanwhile, in tramways it includes several very short sections in remote parts of the city which remain without utter importance to the entire network.

\begin{figure}[h]
\centering\includegraphics[width=1\linewidth]{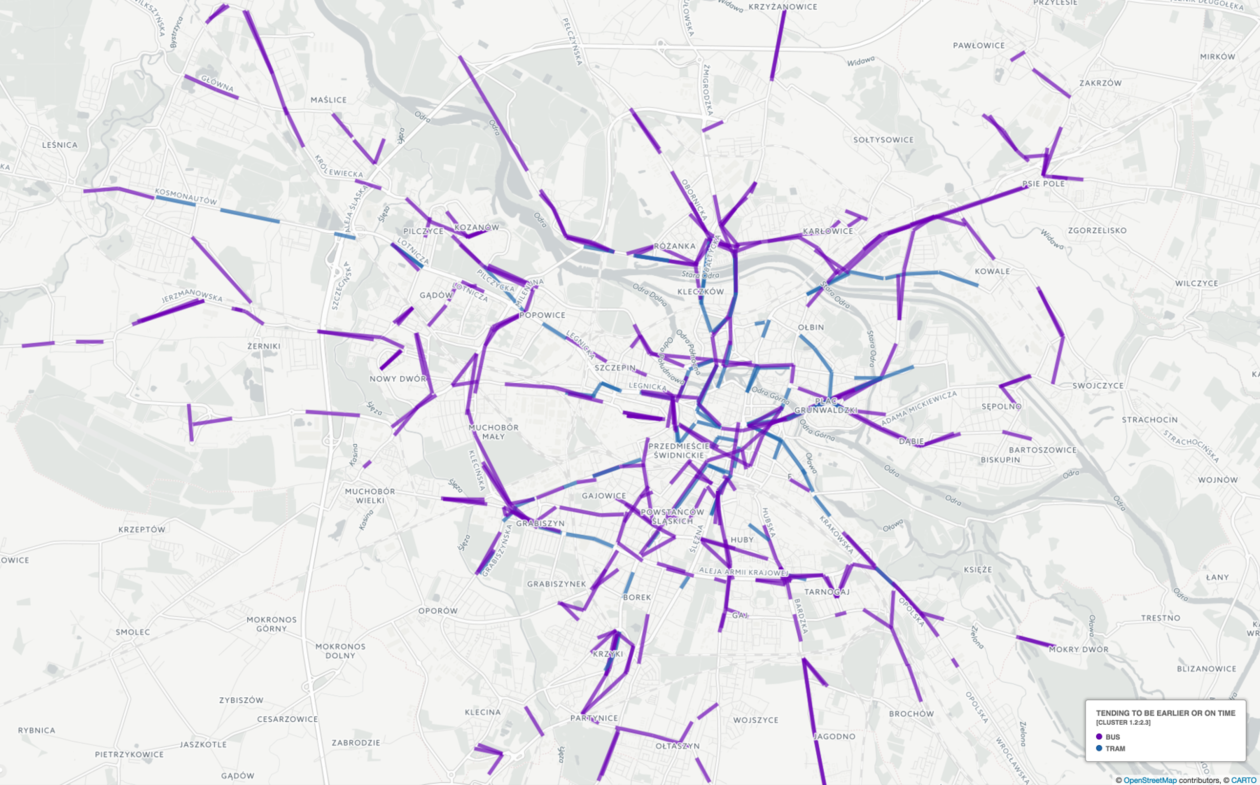}
\caption{Cluster 4 mapped separately, divided by mode of transport (t)ram/(b)us}
\label{fig:cl4_map}
\end{figure}

\subsubsection{Cluster 4: Likely to decrease delay}
It can be clearly seen that both in tramways and buses many timetables include excessive reserves. As this phenomenon occurs at almost all important transit corridors, it seems that timetables should be thoroughly revised. Furthermore, a thorough program should be introduced in order to eliminate significant alterations of travel time between adjacent stops, mostly caused by traffic signals.

\section{Conclusions and Future Work}

We collected and processed large scale AVL data concerning public transport in the city of Wrocław. After processing and official schedule validation 1648 stop to stop edges and related 15 million delay data points were evaluated. We provide a methodology for evaluating delay pattern changes and clustering them into meaningful profiles. Our evaluation provides us with four important profiles of stop to stop edges in the evaluated public transport system: stop pairs that are not impacting the delay significantly, or are likely to cause increase of delay, or are likely to cause strong decrease of delay, or are likely to cause small decrease of delay. We also make a systematic note that delays stem from accumulating small delays over multiple edges rather than from single edges causing vast delays. We provide expert insight into reasons of delay change patterns in each of the clasters based on their spatial distribution and mode of transport.

The area of future work is extensive. We plan to perform extensive analysis of deeper cuts of the dendrogram and subclusters which contain edges with extreme delay change patterns. It is also interesting to move from the cluster analysis to the network analysis and evaluate how the delay change profiles are distributed over paths of lines operating in the city. We expect to find patterns of delay change profiles in lines depending on their spatial and mode of transport characteristics. 

\bibliographystyle{splncs03}
\bibliography{biblio}

\end{document}